\documentclass[apsrev,prl,twocolumn,superscriptaddress]{revtex4-2}
\usepackage{amssymb,amsfonts,amsmath}
\usepackage{adjustbox}
\usepackage{hyperref}
\usepackage{color}

\begin{document}

\title{Elasticity of 2D ferroelectrics across their paraelectric phase transformation}
\author{Joseph E. Roll}
\affiliation{Department of Physics, University of Arkansas, Fayetteville, Arkansas 72701, United States}
\author{John M. Davis}
\affiliation{Department of Physics, University of Arkansas, Fayetteville, Arkansas 72701, United States}
\author{John W. Villanova}
\affiliation{Department of Physics, University of Arkansas, Fayetteville, Arkansas 72701, United States}
\author{Salvador Barraza-Lopez}
\email{sbarraza@uark.edu}
\affiliation{Department of Physics, University of Arkansas, Fayetteville, Arkansas 72701, United States}
\affiliation{MonArk NSF Quantum Foundry, University of Arkansas, Fayetteville, Arkansas 72701, United States}

\begin{abstract}
The mechanical behavior of two-dimensional (2D) materials across 2D phase changes is unknown, and the finite temperature ($T$) elasticity of paradigmatic SnSe monolayers---ferroelectric 2D materials turning paraelectric as their unit cell (u.c.) turns from a rectangle onto a square---is described here in a progressive manner. To begin with, their zero$-T$ {\em elastic energy landscape} gives way to (Boltzmann-like) averages from which the elastic behavior is determined. These estimates are complemented with results from the strain-fluctuation method, which employs the energy landscape or {\em ab initio} molecular dynamics (MD) data. Both approaches capture the coalescence of elastic moduli $\langle C_{11}(T)\rangle=\langle C_{22}(T)\rangle$ due to the structural transformation. The broad evolution and sudden changes of elastic parameters $\langle C_{11}(T)\rangle$, $\langle C_{22}(T)\rangle$, and $\langle C_{12}(T)\rangle$ of these atomically-thin phase-change membranes establishes a heretofore overlooked connection among 2D materials and soft matter.
\end{abstract}
\date{\today}

\maketitle

{\em Introduction.} Zero$-T$ estimates of elastic parameters (sometimes called elastic {\em constants}) lose meaning on materials undergoing phase transitions (transformations) at finite $T$, where elastic behavior is expected to change drastically. For example, zero-$T$ elastic parameters $C_{11}^{(0)}$ and $C_{22}^{(0)}$ have different magnitudes on materials with a rectangular (or orthorhombic in 3D) u.c., but these elastic moduli must turn identical at a critical $T$ ($T_c$) in which the u.c.~turns square (tetragonal, or cubic in 3D).

\begin{figure}[tb]
\begin{center}
\includegraphics[width=0.42\textwidth]{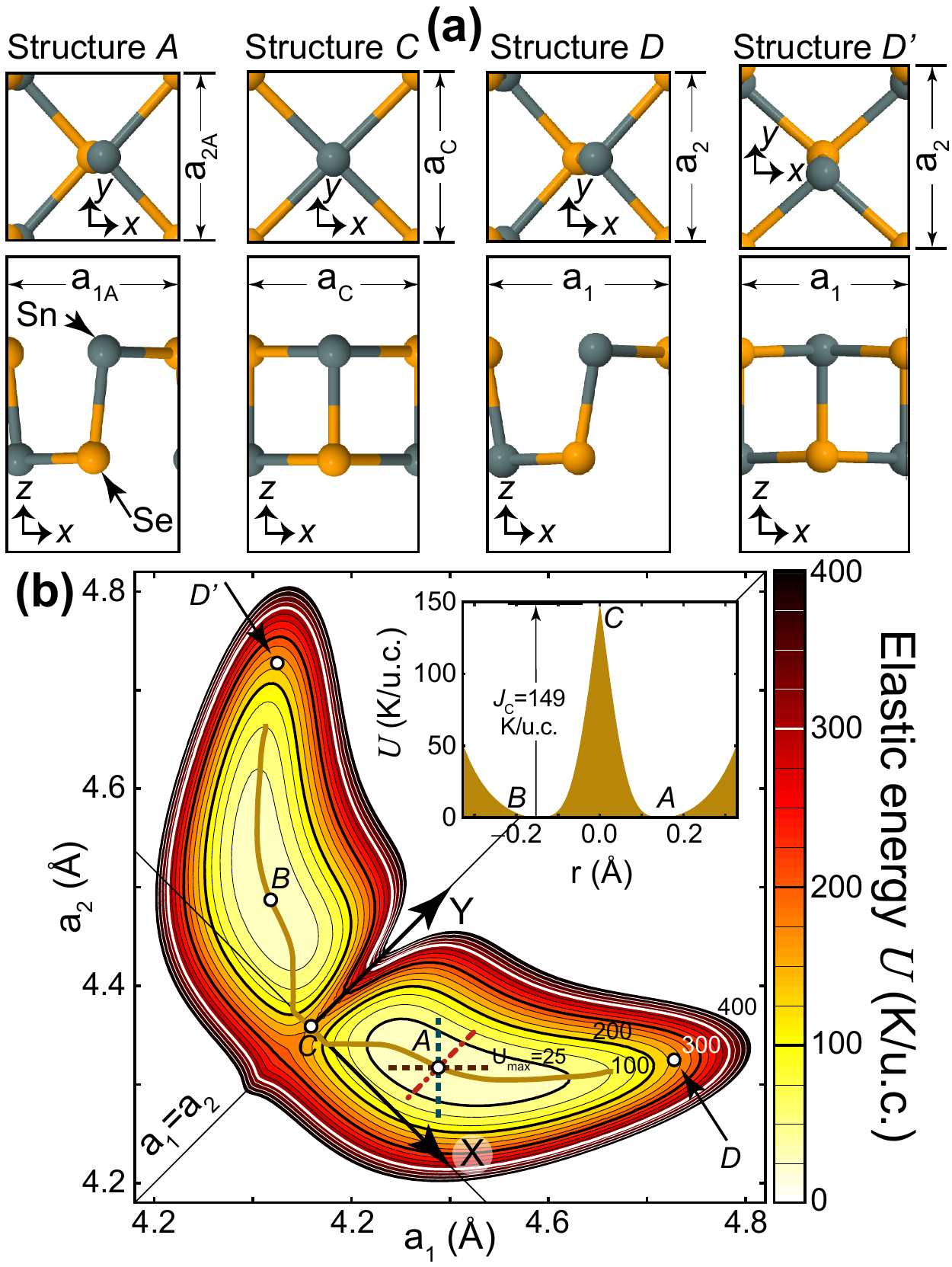}
\caption{(a) SnSe ML u.c.~for points $A$, $C$, $D$, and $D'$ on an analytical $U(a_1,a_2)$ [subplot (b)]. The solid curve connecting points $A$, $C$, and $B$ on subplot (b) is the minimum-energy pathway among the two energy degenerate basins $A$ and $B$, and the inset displays the energy barrier $J_C=U(a_C,a_C)$. Straight lines passing through point $A$ were used to determine zero-$T$ elastic moduli. Some isoenergy lines were drawn, too.}
\label{fig:1}
\end{center}
\end{figure}

Group-IV monochalcogenide monolayers (MLs) are experimentally available \cite{KAICHANGa,KAICHANGb,SNS20} 2D ferroelectrics with a puckered {\em rectangular} u.c.~and a Pnm2$_1$ group symmetry in their zero-$T$ phase, whereby each atom is threefold coordinated \cite{YE17,CUI18,WU18,GUAN20,MEHB16_nano,MEHB16_prl,landscape,VILL20,VILL21,Colloq}.  They display {\em metavalent} bonding \cite{RONN20}, characterized by large atomic effective charges, structural anharmonicity, and significant linear and non-linear optical  responses. Their low-$T$ crystal structure also underpins {\em anisotropic} elasticity \cite{FEI15,GOMES15}. Nevertheless, these 2D materials undergo a firmly  established structural change onto a fivefold coordinated {\em square} structure with P4/nmm symmetry at a critical temperature $T_c$ ranging between 200 and 300 K \cite{KAICHANGa,KAICHANGb,SNS20,landscape,VILL20,Colloq}, at which their properties turn {\em isotropic}. Nothing has been said about the elastic behavior on their P4/nmm phase yet, and approaches based on (i) an analytical form of the zero$-T$ elastic energy landscape \cite{PhysRevB.99.104108}, and (ii) the strain-fluctuation method \cite{strainfluctuation} are deployed to answer this open question here.

{\em Numerical methods.} The elastic energy landscape and MD data were calculated with the {\em SIESTA} DFT code \cite{Martin,siesta} employing an exchange correlation functional with self-consistent van der Waals corrections \cite{soler}. Additional details can be found in Ref.~\cite{landscape}.

{\em Elasticity from elastic energy landscape.} As illustrated on Fig.~\ref{fig:1}(a) for a SnSe ML (a paradigmatic group-IV monochalcogenide ML), a crystal elongated or compressed along two orthogonal directions $a_1$ and $a_2$ with a subsequent structural optimization of atomic positions for a given value of $a_1$ and $a_2$ leads to a zero$-T$ elastic energy $E(a_1,a_2)$ per u.c. The change of energy $U(a_1,a_2)=E(a_1,a_2)-E(a_{1A},a_{2A})$ with respect to a degenerate local minimum energy configuration---labeled $A$ and having coordinates $a_{1A}$ and $a_{2A}$---seen on Fig.~\ref{fig:1}(b) is an {\em elastic energy landscape} \cite{WALES}. To simplify an eventual extraction of partial derivatives, the landscape $U(a_1,a_2)$ in Fig.~\ref{fig:1}(b) is an analytical fit to raw {\em ab initio} data \cite{landscape}. The raw data sets an energy barrier separating the two degenerate minima equal to $J_{C,r}=149.25$ K/u.c., lattice parameters $a_{1A,r}=4.4873$ \AA{}, $a_{2A,r}=4.3264$ \AA{} at the energy minima $A$, and $a_C=4.3590$ \AA{} at for the square u.c.~of lowest energy \cite{landscape,SHIVA19}.

$U(a_1,a_2)$ is mirror symmetric with respect to the $a_1=a_2$ line on Fig.~\ref{fig:1}(b), thus calling for new variables:
\begin{equation}\label{eq:e1}
X=(a_1-a_2)/\sqrt{2}, \text{ and } Y=(a_1+a_2-2a_C)/\sqrt{2}.
\end{equation}
$X=0$ and $Y=0$ at point $C$ (whose coordinates are $a_1=a_2=a_C$) which thus becomes the new origin of coordinates.

The mirror symmetry of the landscape about the $X=0$ line makes $U(X,Y)$ even on $X$, and the following expression was used to fit numerical data \cite{landscape}:
\begin{eqnarray}\label{eq:e2}
U(X,Y)&= J_C + \mathcal{U}_1X^2 + \mathcal{U}_2Y^2 + \mathcal{U}_3YX^2\\
&+ \mathcal{U}_4Y^3 + \mathcal{U}_5X^4 + \mathcal{U}_6Y^4 \nonumber\\
+(\mathcal{U}_7Xe^{-\sqrt{X^2}/g_1}&+\mathcal{U}_8YXe^{-\sqrt{X^2}/g_2})\text{tanh}(100X),\nonumber\
\end{eqnarray}
with parameters and numerical uncertainties provided in Table \ref{ta:ta1}. With the exception of the terms on tanh($100X$)---whose sole purpose is to smooth the cusp observed at the barrier in the numerical data \cite{landscape}; see inset of Fig.~\ref{fig:1}(b)---the elastic energy landscape is a polynomial of order four. The quality of the fitting can be ascertained by noticing that its minima $A$ is located at $(a_{1A},a_{2A})=(4.4896 \text{ \AA} ,4.3173\text{ \AA})$ [or $X_A=0.1218$ \AA, $Y_A=0.0629$ \AA], which is less than 0.25\% different from the raw {\em ab initio} data. One also notices that the saddle point on $U(X,Y)$ (i.e., the minimum energy barrier separating the two ground states $A$ and $B$) occurs exactly at point $a_C$ as determined in the raw data, and that $U(X_A,Y_A)=0.0245$ K/u.c., leading to an energy barrier of 148.9755 K/u.c.~which is only 0.2745 K/u.c.~smaller than the one seen from the raw data.

Zero$-T$ elastic moduli $C_{11}^{(0)}$, $C_{22}^{(0)}$ and $C_{12}^{(0)}$ are customarily obtained  by fitting $U(a_1,a_2)$ to parabolas \cite{FEI15,GOMES15}:
\begin{equation}\label{eq:e3}
U\simeq \mathfrak{U}=\frac{1}{2}\boldsymbol{\epsilon}^T\mathcal{C}^{(0)}\boldsymbol{\epsilon}=
\frac{C_{11}^{(0)}\epsilon_1^2}{2}+\frac{C_{22}^{(0)}\epsilon_2^2}{2}+C_{12}^{(0)}\epsilon_1\epsilon_2,
\end{equation}
where {\em strain} coordinates $\boldsymbol{\epsilon}=(\epsilon_1,\epsilon_2)^T$ with
\begin{equation}\label{eq:e4}
\epsilon_1=(a_1-a_{1A})/a_{1A}, \text{ } \epsilon_2=(a_2-a_{2A})/a_{2A},
\end{equation}
were employed. We recall that $a_{1A}$ and $a_{2A}$ in Eqn.~\eqref{eq:e4} are zero$-T$ equilibrium lattice parameters defining point $A$ in the elastic energy landscape.

\begin{figure}[tb]
\begin{center}
\includegraphics[width=0.47\textwidth]{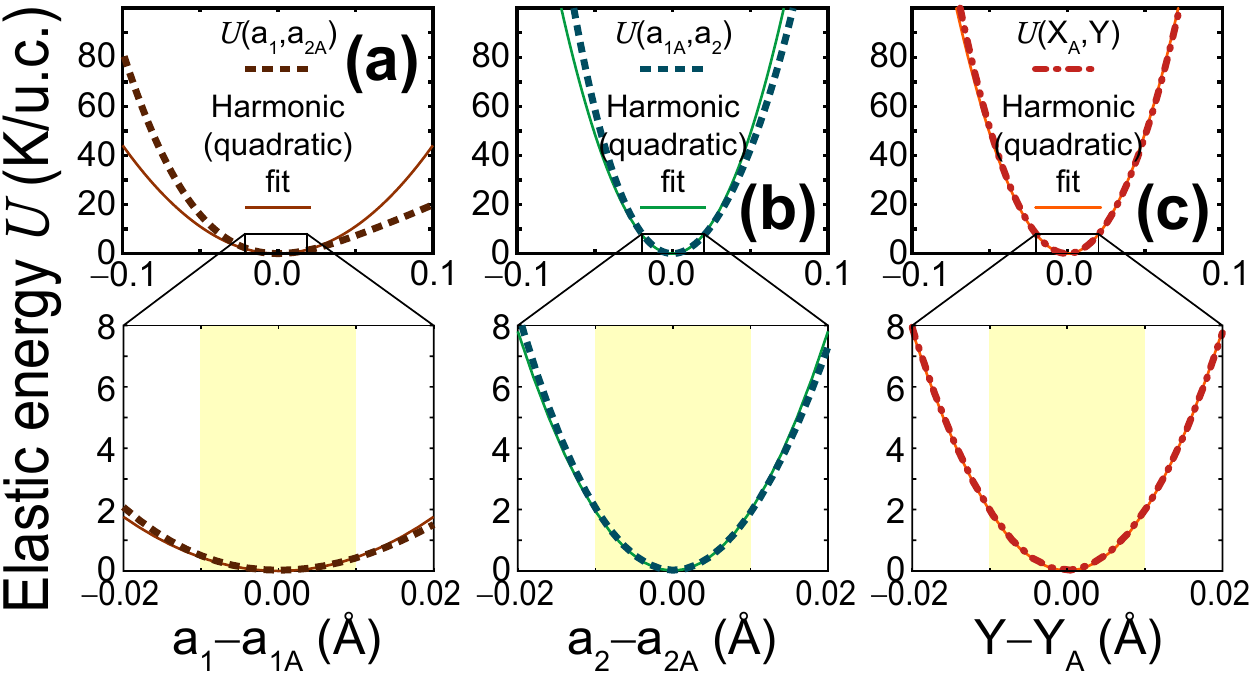}
\caption{Cuts of $U$ along straight lines passing through point $A$ on Fig.~\ref{fig:1}(b), and harmonic (i.e., quadratic) fits---thinner solid curves obtained within the shaded regions on the zoom-in plots---from which $C_{11}^{(0)}$, $C_{22}^{(0)}$, and $C_{12}^{(0)}$ were extracted.}
\label{fig:2}
\end{center}
\end{figure}

$\mathcal{C}^{(0)}$ is the {\em harmonic approximation} to the elasticity tensor, and $\mathfrak{U}$ is the {\em harmonic approximation to} $U$. As acutely seen in Fig.~\ref{fig:2}(a), the prescription within Eqn.~\eqref{eq:e3} neglects the strong anharmonicity of group-IV monochalcogenide MLs {\em by definition}. Further, given that elastic moduli are thermodynamical averages after all, {\em such approach misses a finite-$T$ understanding of elasticity altogether}.

\begin{table}
\caption{Fitting parameters for $U(X,Y)$. $J_C=149$ K/u.c.\label{ta:ta1}}
\begin{center}
\begin{tabular}{c|c||c|c}
\hline
\hline	
$\mathcal{U}_1$ & $-$3660.5 $\pm$ 12.3\% K/\AA$^2$ & $\mathcal{U}_2$ & 24849    $\pm$  4.3\% K/\AA$^2$\\
$\mathcal{U}_3$ & $-$109410 $\pm$ 8.2\%  K/\AA$^3$ & $\mathcal{U}_4$ & $-$42945 $\pm$ 21.2\% K/\AA$^3$\\
$\mathcal{U}_5$ & 188100    $\pm$ 9.2\%  K/\AA$^4$ & $\mathcal{U}_6$ & 114840   $\pm$ 43.4\% K/\AA$^4$\\
$\mathcal{U}_7$ & $-$3568.5 $\pm$ 13.4\% K/\AA     & $\mathcal{U}_8$ & $-$88140 $\pm$ 12.4\% K/\AA$^2$\\
$g_1$ & 0.0583 $\pm$ 9.3\% \AA                     & $g_2$ & 0.0536 $\pm$ 8.1\% \AA \\
\hline
\hline
\end{tabular}
\end{center}
\end{table}

$U(X,Y)$ leads to zero$-T$ elastic moduli consistent with prior work \cite{FEI15,GOMES15}: Eqn.~\eqref{eq:e2} is calculated along three straight lines [($a_1,a_{2A}$), ($a_{1A},a_2$), and ($X_A,Y$), corresponding to the brown (horizontal), green (vertical), and red (at $45^{\circ}$) straight lines passing through point $A$ on Fig.~\ref{fig:1}(b), respectively] and
Eqn.~\eqref{eq:e3} is fitted against the parabolas displayed on Fig.~\ref{fig:2}. $C_{ij}^{(0)}$ are listed in Table \ref{ta:ta2} ($i,j=1,2$). Discrepancies with previous results (such as the smaller magnitude of $C_{11}^{(0)}$ and the slightly larger value of $C_{12}^{(0)}$ than $C_{11}^{(0)}$ here) are due to the use of different computational tools and exchange-correlation functionals in {\em ab initio} calculations. The softer $C_{11}^{(0)}$ here leads to a smaller $T_c$ when contrasted to results using the numerical methods of Refs.~\cite{GOMES15,FEI15}; see Refs.~\cite{VILL20} and \cite{Colloq} for a discussion.

\begin{table}
\caption{Zero$-T$ in-plane elastic moduli (N/m).}
\label{ta:ta2}
\begin{center}
\begin{tabular}{c|c|c}
\hline
\hline
Elastic modulus & Prior work & This work \\
\hline
\hline
$C_{11}^{(0)}$ & 19.9 \cite{FEI15}, 19.2 \cite{GOMES15}  &  12.7 \\
\hline
$C_{22}^{(0)}$ & 44.5 \cite{FEI15}, 40.1 \cite{GOMES15}  &  51.8 \\
\hline
$C_{12}^{(0)}$ & 18.6 \cite{FEI15}, 16.0 \cite{GOMES15}  &  20.9 \\
\hline
\hline
\end{tabular}
\end{center}
\end{table}

\begin{figure*}[tb]
\begin{center}
\includegraphics[width=0.94\textwidth]{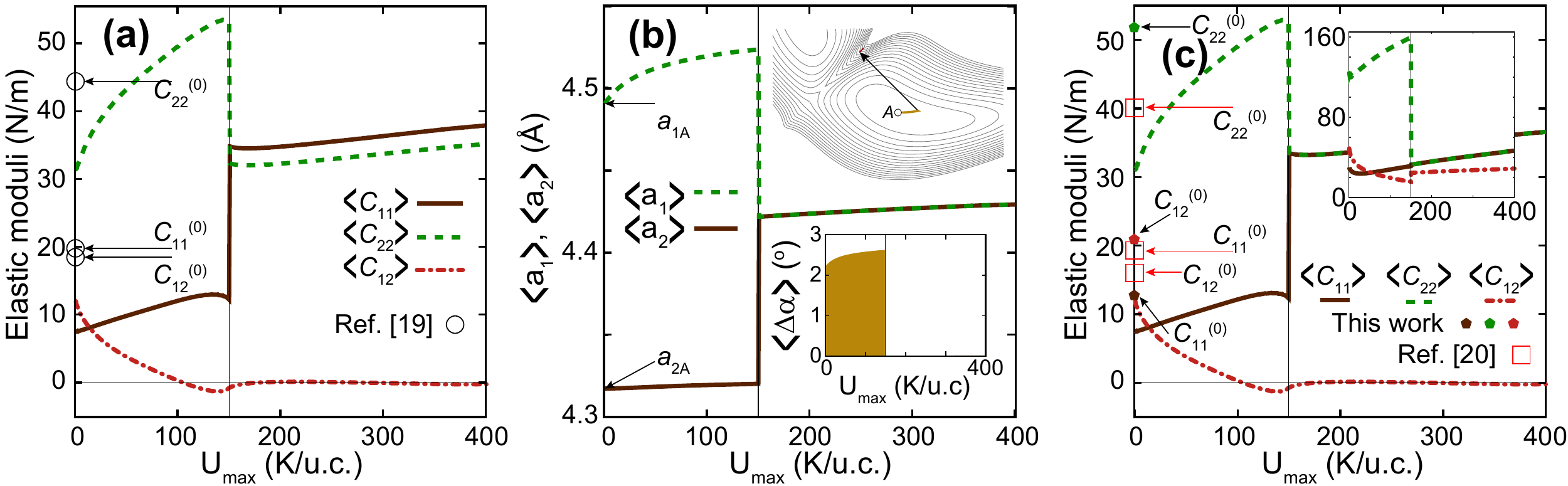}
\caption{(a) Elastic moduli as a function of $U_{max}-$isovalue, setting the strain with respect to $a_{1A}$ and $a_{2A}$ at zero$-T$. (b) Average lattice constants {\em versus} $U_{max}$: the u.c.~turns from a rectangle ($\langle a_1\rangle >\langle a_2\rangle$) onto a square ($\langle a_1\rangle=\langle a_2\rangle$) when $U_{max}\ge J_C$. Insets: evolution of $\langle \Delta\alpha \rangle$ and of the point $(\langle a_1\rangle,\langle a_2\rangle)$ {\em versus} $U_{max}$. (c) Elastic moduli {\em versus} $U_{max}$, setting the strain with respect to $\langle a_{1}\rangle$ and $\langle a_{2}\rangle$ as obtained at subplot (b): $\langle C_{11}\rangle=\langle C_{22}\rangle$ are now identical past $J_C$. Inset: elastic moduli within the strain-fluctuation method. $C_{11}^{(0)}$, $C_{22}^{(0)}$, and $C_{12}^{(0)}$ as estimated by us and others are shown in subplots (a) and (c).}
\label{fig:3}
\end{center}
\end{figure*}

To go beyond the zero$-T$ paradigm, we make use of $U(X,Y)$ to determine elastic behavior next. A function of $X$ and $Y$ has an expectation value within the elastic energy landscape $\langle f(U_{max})\rangle$  as an {\em average over classically accessible states} \cite{PhysRevB.99.104108}:
\begin{equation}\label{eq:e5}
\langle f(U_{max})\rangle=\frac{\oint e^{-U(X,Y)/U_{max}} f(X,Y) dXdY}{\oint e^{-U(X,Y)/U_{max}} dXdY},
\end{equation}
with $dXdY$ an area element within the confines of a isoenergy contour $U_{max}$ around structure $A$, like those seen on Fig.~\ref{fig:1}(b).

Within this paradigm, $U(X,Y)$ is a classical potential energy profile, and a set of accessible {\em crystalline configurations} lies within isoenergy confines. [$U_{max}$ is the largest kinetic energy of a hypothetical particle in the landscape, and is thus indirectly linked to $T$ that way.] For example, sampled u.c.s will all have $a_1>a_2$ when the $U_{max}$ isoenergy curve is smaller than $J_C$. This is, the sampled structures will all be {\em ferroelectric}, having an in-plane polarization along the $x-$direction \cite{landscape}; see structure $A$ on Fig.~\ref{fig:1}(a). When $U_{max}\ge J_C$ nevertheless, the {\em average structure} encompasses minima $A$ and $B$ yielding $a_1=a_2$, and it thus is a square. The fact that $a_1=a_2$ on average when $U_{max}\ge J_C$ is illustrated by structures $D$ and $D'$ on Fig.~\ref{fig:1}(b), which have $x-$ and $y-$coordinates swapped. In this sense, the averaging among {\em crystalline configurations} within the energy landscape  up to an energy $U_{max}$ achieves an effect similar to $T$: a transformation whereby the average u.c.~turns from a rectangle onto a square. A caveat to this model is that it is based on {\em averaging over independent crystalline} u.c.s, while 2D structural transformations in 2D are driven by disorder \cite{MEHB16_nano,MEHB16_prl}.

Energy average values for $C_{ij}$ are determined by \cite{PhysRevB.99.104108}:
\begin{eqnarray}\label{eq:e6}
\langle C_{ij}(U_{max})\rangle =&k_B\bigg\{\bigg\langle\frac{\partial^2u}{\partial\epsilon_i\partial\epsilon_j}\bigg\rangle\\
-\frac{1}{U_{max}}\bigg[\bigg\langle \mathcal{A}\frac{\partial u}{\partial\epsilon_i}\frac{\partial u}{\partial\epsilon_j}\bigg\rangle-&\langle \mathcal{A}\rangle \bigg\langle \frac{\partial u}{\partial\epsilon_i}\bigg\rangle \bigg\langle \frac{\partial u}{\partial\epsilon_j}\bigg\rangle \bigg]\bigg\},\nonumber	
\end{eqnarray}
with $k_B$ Boltzmann's constant, $\mathcal{A}=a_1a_2$, and $u=U/\mathcal{A}$.

\begin{figure*}[tb]
\begin{center}
\includegraphics[width=0.94\textwidth]{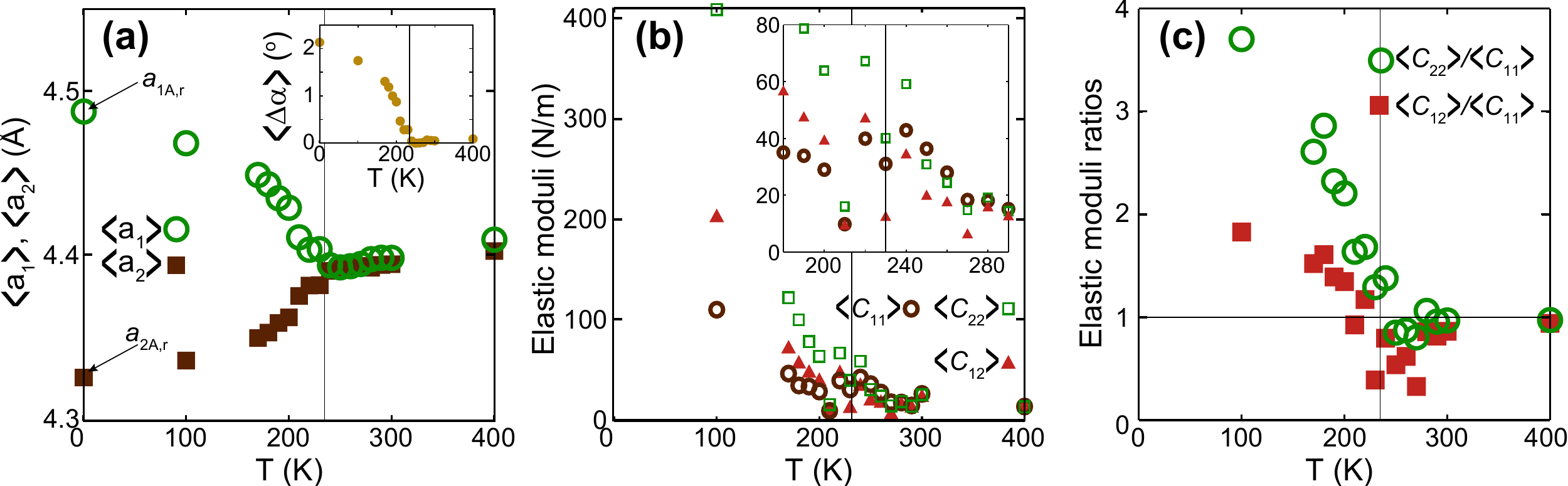}
\caption{(a) $\langle a_1\rangle$ and $\langle a_2 \rangle$ {\em versus} $T$ on MD calculations. Inset: $\langle\Delta \alpha\rangle$. (b) Elastic moduli from the strain-fluctuation method and $\langle \epsilon_i\epsilon_j\rangle$, $\langle \epsilon_i\rangle$ ($i=1,2$) determined from MD: see the coalescence of $\langle C_{11}\rangle$ and $\langle C_{22}\rangle$ past $T_C$, and the similar magnitudes of $\langle C_{ij}\rangle$ to those seen at the inset of Fig.~\ref{fig:3}(c). (c) Ratio among the elastic moduli displayed in subplot (b).}
\label{fig:4}
\end{center}
\end{figure*}

Eqn.~\eqref{eq:e6} was evaluated numerically for energy isovalues $U_{max}$ starting at 1K/u.c and up to 400 K/u.c. [Fig.~\ref{fig:3}(a)]. Near $U_{max}=0$, the averaging procedure yields elastic moduli smaller than those listed in Table \ref{ta:ta2}. Within this method, $\langle C_{12}\rangle$ quickly decays to a nearly zero value and it becomes negative (for an auxetic behavior). On the other hand, $\langle C_{22}\rangle >\langle C_{11}\rangle$ by a factor in between 3 and 5 for energies up to $U_{max}=J_C$, at which a sharp change occurs whereby $\langle C_{22}\rangle \approx \langle C_{11}\rangle$.

The fact that $\langle C_{22}\rangle \ne \langle C_{11}\rangle$ for isovalues $U_{max}\ge J_C$, in which the average structure already turned isotropic [see Fig.~\ref{fig:3}(b)], represents an inaccuracy of the approach in Ref.~\cite{PhysRevB.99.104108}. It originates from the fact that strain was written out with respect to the zero$-T$ ground state structure ($a_{1A},a_{2A}$) in Eqn.~\eqref{eq:e4}. Experimentally, strain at finite$-T$ is measured with respect to a structure in {\em thermal equilibrium}, calling for a calculation of elastic moduli in which average values of $a_{1}$ and $a_{2}$ are employed. Strain is then redefined as:
\begin{equation}\label{eq:e7}
\epsilon_1=\frac{a_1-\langle a_{1}\rangle}{\langle a_{1}\rangle},\text{ and }\epsilon_2=\frac{a_2-\langle a_{2}\rangle}{\langle a_{2}\rangle},
\end{equation}
which is still valid at zero$-T$ in which $\langle a_{i}\rangle=a_{iA}$ ($i=1,2$) The resulting elastic parameters are shown in Fig.~\ref{fig:3}(c). Now, $\langle C_{22}\rangle = \langle C_{11}\rangle$ for isovalues $U_{max}\ge J_C$. The use of Eqn.~\eqref{eq:e7} instead of Eqn.~\eqref{eq:e4} is  thus a correction to our previous method \cite{PhysRevB.99.104108}.

$\langle C_{12} \rangle$ is the softest elastic modulus on this model. On the other hand, $\langle C_{11}\rangle$ hardens significantly at the transition ($U_{max}=J_C$), while $\langle C_{22}\rangle$ suddenly softens at $U_{max}=J_C$. According to Fig.~\ref{fig:3}, a SnSe ML is much softer than graphene, for which $C_{11}^{(0)}=C_{22}^{(0)}=336$ N/m, and $C_{12}^{(0)}=75$ N/m (see Ref.~\cite{elasticGraphene}, and multiply by half of Bernal graphite's unit cell thickness $\simeq$3.4 \AA).

We propose---by direct comparison among $J_C$ and $T_C$ from numerical calculations \cite{VILL20}---a linear correspondence among these two variables ($T\propto 1.42 U_{max}$) for this material, such that $T_C=212$ K, and finite$-T$ elastic behavior can be extracted from Fig.~\ref{fig:3} at a low computational cost.

{\em Elasticity from the strain-fluctuation method.} We next employ the strain-fluctuation method to determine the elastic moduli. The expression to work with is \cite{strainfluctuation}:
\begin{equation}\label{eq:e8}
\langle C^{-1}{}_{ij}\rangle=\frac{\langle\mathcal{A}\rangle}{k_BT}\left(\langle \epsilon_i\epsilon_j\rangle- \langle \epsilon_i\rangle \langle\epsilon_j\rangle\right),
\end{equation}
which is less convoluted than Eqn.~\eqref{eq:e6}, and also amenable for MD input.

Computed using $U(X,Y)$, $\langle \epsilon_i\rangle=\frac{\langle a_i -\langle a_{i}\rangle\rangle}{\langle a_{i}\rangle}=\frac{\langle a_i\rangle -\langle a_{i}\rangle}{\langle a_{i}\rangle}=0$ ($i=1,2$) for additional simplification, and $\langle C_{ij}\rangle$ ($i,j=1,2$) are displayed as an inset on Fig.~\ref{fig:3}(c). One notes that $\langle C_{ij}\rangle >0$ now, so that auxetic behavior cannot be confirmed within the strain-fluctuation method. A second point to notice is that $\langle C_{22}\rangle $ now becomes three times larger than its biggest magnitude obtained using Eqn.~\ref{eq:e6}. For $U_{max}<J_C$, $\langle C_{11}\rangle$ is about twice as large than its magnitude from Eqn.~\ref{eq:e6}, too. $\langle C_{11}\rangle =\langle C_{22}\rangle$ for $U_{max}\ge J_C$, with a magnitude now comparable to that obtained from Eqn.~\ref{eq:e6}. The two takeouts from the strain-fluctuation approach [inset on Fig.~\ref{fig:3}(c)] are that $\langle C_{22}\rangle$ is much larger than its estimate using partial derivatives of $U(X,Y)$, and that $\langle C_{21}\rangle$ remains larger than zero.

The Pnm2$_1$ to P4/nmm structural transformation is signaled by a collapse of the rhombic distortion angle $\langle \Delta \alpha \rangle$ [related to $a_1$ and $a_2$ as $\langle \Delta \alpha \rangle=\left(\langle \frac{a_1}{a_2}\rangle-1\right)\frac{180^{\circ}}{\pi}$] to a zero value \cite{KAICHANGa,landscape}. As seen at an inset on Fig.~\ref{fig:3}(b), $U(X,Y)$ does yield the required collapse of $\langle \Delta \alpha \rangle$, but it does not display a gradual decrease with a critical exponent of 1/3 \cite{KAICHANGa,landscape} as the inset on Fig.~\ref{fig:4}(a)---obtained from MD---does. This is so because $U(X,Y)$ makes $\langle a_1\rangle$ plow to larger values while $\langle a_2 \rangle$ remains relatively unchanged up to $U_{max}=J_C$, when both lattice parameters change {\em discontinuously} onto an identical value [see Fig.~\ref{fig:3}(b) and its upper inset].

And thus, while an estimation of elastic properties based on $U(X,Y)$ [using either Eqn.~\eqref{eq:e6}, or Eqn.~\eqref{eq:e8}] is relatively inexpensive, MD data was also utilized to estimate $\langle C_{11}\rangle$, $\langle C_{22}\rangle$, and $\langle C_{12}\rangle$ within the strain-fluctuation approach. Briefly, 16$\times$16 supercells containing 1024 atoms were employed on NPT {\em ab initio} MD calculations for sixteen different $T$s in between 100 and 400 K. 20,000 individual  timesteps with a 1.5 fs resolution were obtained for any given $T$. Thermal averages were obtained for times above 5 ps to allow for proper thermalization. In this approach, $\epsilon_i=\frac{1}{2}\left[(\langle h\rangle^{-1T} h^T h \langle h\rangle^{-1})_{ii}-1 \right]$ ($i=1,2$) \cite{strainfluctuation}. $h=(\mathbf{a}_1,\mathbf{a}_2)$, and $\langle h\rangle =(\langle\mathbf{a}_1\rangle,\langle\mathbf{a}_2\rangle)$ are $2\times 2$ matrices containing the in-plane magnitudes of supercell lattice vectors $\mathbf{a}_1$ and $\mathbf{a}_2$, which are written in column form. The matrix $h$ contains the in-plane superlattice constants for one MD step, and $\langle h\rangle$ is its average over the available MD steps past thermalization. Here, $\langle \mathcal A\rangle$ is replaced by the supercell's area thermal average.

The results, shown in Figs.~\ref{fig:4}(b) and \ref{fig:4}(c), indicate a magnitude of $\langle C_{22}\rangle$ comparable with that of graphene at 100 K \cite{elasticGraphene}, but a softer magnitude of $\langle C_{11}\rangle$ that is four times smaller, as it is expected due to the SnSe ML's anisotropy. All elastic constants then decrease, in a manner similar to that seen at the inset of Fig.~\ref{fig:3}(c). $\langle C_{ii}\rangle$  ($i=1,2$) turn similar despite of method employed at energies/temperatures above the transition.

{\em Conclusion. } The finite$-T$ elastic behavior of a paradigmatic 2D ferroelectric was estimated from second-order partial derivatives of the energy on their zero$-T$ elastic energy landscape, and following the prescriptions of the strain-fluctuation method as well. Within the later method, average strain was introduced utilizing either the elastic energy landscape, or dedicated {\em ab initio} MD data. Despite of method, $\langle C_{11}\rangle$ are shown to coalesce past the transition energy $J_C$ or temperature $T_C$, and the elastic moduli turns much softer than that determined on graphene. The results contained here thus show how to understand the finite-$T$ elastic behavior of 2D materials undergoing two-dimensional transformations.

\begin{acknowledgments}
The authors acknowledge Dr.~P.~Kumar for insightful conversations, as well as support from the U.S.~Department of Energy (J.W.V.~was funded by Award DE-SC0016139, and S.B.L.~by Award DE-SC0022120).
\end{acknowledgments}


\end{document}